\documentclass{article}

    \PassOptionsToPackage{numbers}{natbib}
    \usepackage[preprint]{neurips_2019}

\usepackage[utf8]{inputenc} % allow utf-8 input
\usepackage[T1]{fontenc}    % use 8-bit T1 fonts
\usepackage{hyperref}       % hyperlinks
\usepackage{url}            % simple URL typesetting
\usepackage{booktabs}       % professional-quality tables
\usepackage{amsfonts}       % blackboard math symbols
\usepackage{nicefrac}       % compact symbols for 1/2, etc.
\usepackage{microtype}      % microtypography

\usepackage{natbib}
\title{Algorithmic Injustices: Towards a Relational Ethics}

\author{%
  Abeba Birhane, Fred Cummins \\
  School of Computer Science, University College Dublin, Ireland\\
  \texttt{abeba.birhane@ucdconnect.ie,fred.cummins@ucd.ie }
}

\begin{document}

\maketitle

% Example for abeba for quotaing
% ``cold data``.

% \begin{abstract}
%   The abstract paragraph should be indented \nicefrac{1}{2}~inch (3~picas) on
%   both the left- and right-hand margins. Use 10~point type, with a vertical
%   spacing (leading) of 11~points.  The word \textbf{Abstract} must be centered,
%   bold, and in point size 12. Two line spaces precede the abstract. The abstract
%   must be limited to one paragraph.
% \end{abstract}
\vspace{-0.1cm}
\section{Introduction}
It has become trivial to point out how decision-making processes in various social, political and economical sphere are assisted by automated systems. Improved efficiency, the hallmark of these systems, drives the mass scale integration of automated systems into daily life. However, as a robust body of research in the area of algorithmic injustice shows~\cite{o2016weapons, ananny2018seeing, caliskan2017semantics, benjamin2019race}, algorithmic tools embed and perpetuate societal and historical biases and injustice. In particular, a persistent recurring trend within the literature indicates that society’s most vulnerable are disproportionally impacted. When algorithmic injustice and bias is brought to the fore, most of the solutions on offer 1) revolve around technical solutions and 2) do not focus centre disproportionally impacted groups. This paper zooms out and draws the bigger picture. It 1) argues that concerns surrounding algorithmic decision making and algorithmic injustice require fundamental rethinking above and beyond technical solutions, and 2) outlines a way forward in a manner that centres vulnerable groups through the lens of relational ethics. 
\section{The Problem}
\vspace{-0.1cm}
Devices that proliferate public and private space (e.g., phones, cameras, IoT sensors), produce mass flow of data that are fed into machine learning systems. Dynamic, pattern-based abstractions provide these machine learning systems with actionable indices of ``who we are'' and ``how we might behave''~\cite{zuboff2019age, cheney2018we}. When these machine learning systems that infer and predict individual behaviour and action, based on superficial extrapolations, are deployed into the social world, various unintended problems arise.~These systems ‘pick up’ social and historical stereotypes rather than any deep fundamental causal explanations. In the process, individuals and groups, often at the margins of society that fail to fit stereotypical boxes, suffer the undesirable consequences. Various findings illustrates this: bias in detecting skin tones in pedestrians~\cite{wilson2019predictive}; bias in predictive policing systems~\cite{richardson2019dirty}; gender bias and discrimination in the display of STEM career ads~\cite{lambrecht2019algorithmic}; racial bias in recidivism algorithms~\cite{angwin2016machine}; bias in the politics of search engines~\cite{introna2000shaping}; bias and discrimination in medicine~\cite{ferryman2018fairness}; and bias in hiring~\cite{ajunwa2016hiring}, to mention but a few. 
As algorithmic decision-making is deployed across the social sphere in socially contested contexts such as ``acceptable'' behaviour, ``ill'' health, and ``normal'' body type, algorithms make decisions with moral implications, not just value-free categorization tasks, for instance. Despite this fact, their operations are seen as technical which is often taken to be ``value-free'' and ``a-moral''~\cite{mcquillan2018data}. Unfair and discriminatory outcomes, as a result, are treated as side-effects that have technical solutions rather than problems with fundamentally deep-rooted underlying assumptions. Research in algorithmic fairness almost exclusively focuses on technical ``implementable'' solutions. Although these solutions are necessary, they can be superficial, hiding bias (not removing it) at best, and thus giving the illusion that the problem has been solved~\cite{gonen2019lipstick}. 
\section{Motivation and Contribution }
The first move towards Relational ethics is an acknowledgement that when we are dealing with data~(especially those that measure, classify, and/or analyse the social, cultural, and political realm), we are necessarily dealing with human culture, value, and meaning. Consequently, any tool we create and deploy are intimately tied with the nature of being and morality – questions that necessarily require us to look beyond ``purely technical solutions''. Relational perspectives \cite{di2018linguistic, holquist2003dialogism}  encourage us to pull the rug from under and examine fundamental questions and unstated assumptions. To this end, this paper proposes to go beyond ``technical solutions'' and emphasizes the need for critical examination of deeply rooted, unexamined assumptions. It explores deeper philosophical questions that need to be addressed as part of developing and deploying ethical algorithmic systems. These include the need to treat terms such as ``bias'', ``justice'', ``fairness'', and ``ethics'' as dynamic, moving, and contingent concepts that never emerge in a social and political vacuum, which consequently require more than technical and methodological solutions. 
\section{Towards a Relational Ethics }
\subsection{Centering the disproportionally impacted}
\vspace{-0.1cm}
For any individual person, group, or situation, algorithmic classifications and predictions give an advantage or they give suffering. Certain patterns are made visible and objectified while other types are erased. Some identities (and not others) are recognised as a pedestrian, or fit for a STEM career, or in need of medical care. Some are identified as criminal while others are ignored by algorithmic machines altogether. The central recurring theme in algorithmic injustice literature remains that individuals and groups that are at the margin of society are disproportionally impacted. Relational ethics asks that for any solution that we seek, the starting point be the individuals and groups that are impacted the most. This means we seek to centre those disproportionally impacted and not solutions that benefit the majority. 

\subsection{Prioritization of understanding over prediction}
\vspace{-0.1cm}
Much work within developing or implementing algorithmic systems is focused on maximizing prediction and accuracy with very little effort, if any at all, going towards investigating deeper understandings. Within prison systems, for example, the focus is on finding patterns on criminal activities with the aim of developing systems that predict future crimes. Relational ethics emphasizes that we shift the focus towards gaining deeper contextual understandings prior to developing predictive models. In the case of the prison system, for example, we ask why we find the patterns that we do, which in turn calls for contextual and historical antecedents, instead of using our findings as inputs towards predictive systems. 

\subsection{Algorithms as more than tools that create and sustains certain social order 
}
\vspace{-0.1cm}
The persistent perception of algorithmic tools as merely a method rather than a way of defining, classifying, and organizing the person and furthermore the social world, contributes to why ``technical solutions'' to algorithmic bias and unfairness seem logical.~Classification, categorization, and pattern recognition are assumed to be technical and methodological problems that we need to ``solve'' rather than practices that create certain social order~\cite{bowker2000sorting}.~In reality, algorithmic classifications, predictions, and standards that seem ordinarily invisible in our daily lives, create and sustain the social and moral order. The use of hiring and policing algorithms, for example, requires the simplification of contested and complex terms such as ``successful'' and ``criminality''.~Such seemingly banal tasks carry grave consequences for certain individuals and groups, and subsequently reconfiguring socially acceptable norms~\cite{fletcher2018diversity}. Relational ethics encourages us to view the task of developing and deploying algorithmic tools as part of the practice of creating and reinforcing certain types of norms and social~order. 

\subsection{Bias, fairness, and justice are moving targets}
\vspace{-0.1cm}
Neither people, nor the environment, are static; what society deems fair and ethical changes over time. The concept of fairness and ethical practice is, therefore, a moving target and not something that can have a final answer or can be ``solved'' once and for all. It is possible that what is considered ethical currently and within certain domains for certain societies will not be received similarly at a different time, in another domain, or for a different society~\cite{verbeek2008morality}. Adopting relational ethics means that we view our understandings, proposed solutions, and definitions of ``bias'', ``fairness'' and ``ethics'' as partially-open. This partial openness allows for revision and reiteration in accordance with the dynamic development of such challenges.

\subsection*{Acknowledgments}
This work is supported, in part, by Science Foundation Ireland grant 13/RC/2094.

\small
  \bibliographystyle{abbrv}

\bibliography{references.bib}

% \addbibresource{reference.bib}

\end{document}